\pdfoutput=1
\documentclass{article}
\usepackage{spconf,amsmath,amssymb,graphicx,booktabs,balance,hyperref}
\hypersetup{colorlinks=true,urlcolor=blue,linkcolor=black,citecolor=black}
\title{Game Sound-Effect Completion with Event-Level Transformation Hints}
\name{Xinrui Jiang$^1$ \qquad Heng Yu$^2$}
\address{$^1$Department of Electrical Engineering, $^2$Department of Computer Science\\
Stanford University \quad \texttt{\{jiangxr,yuheng\}@stanford.edu}}
\begin{document}
\ninept
\maketitle
\begin{abstract}
Creating sound effects for a new game-character skin requires a distinct acoustic identity while preserving gameplay-event roles. The challenge is to complete a coherent set of related sounds whose required degrees of redesign differ. We formulate this task as completion conditioned on base-skin audio, completed target assets, and a textual design description. We develop a pipeline to collect, process, and align corresponding events across League of Legends skins. Building on Stable Audio 3's pretrained audio prior, we fine-tune a latent inpainting model to jointly complete missing events. A signed soft retention mask encodes available audio and an adjustable transformation hint for each missing event, specifying the requested balance between retention and redesign. Experiments on held-out skins show improved reconstruction over the evaluated general-purpose audio editors. Target-derived hints further improve paired similarity, with three-level hints retaining most of \mbox{the benefit of continuous guidance}.
\end{abstract}

\section{Introduction}
\label{sec:intro}
Sound effects (SFX) communicate game events, provide feedback on player actions, and contribute to immersion. In character-based games such as \textit{League of Legends}, a cosmetic skin creates a sound-design problem: an ability retains its gameplay role while its presentation may change in material, timbre, or decorative elements. Plausible audio in isolation is insufficient; the assets must fit their event roles and the target design. A recent study of sound-design practitioners, including game-audio practitioners, finds demand for task-specific assistance and parametric control~\cite{garcia2026workflows}. This motivates a completion workflow that retains the designer's existing decisions and exposes controls over the remaining work.

Available design information does not uniquely specify a sound. Artwork, narrative descriptions, and animations suggest a theme, but leave attacks, textures, and acoustic layers undetermined. Base-skin audio supplies a concrete event reference; already completed target-skin assets supply audible examples of the intended design. Even with these inputs, the required degree of change varies by event: some sounds can be retained, whereas others need substantial redesign. We represent this decision with a scalar \emph{transformation hint} for each unknown event.

Studying this workflow requires structured data. Gameplay recordings mix ability effects with other sounds, complicating isolated event correspondence. Our pipeline extracts game-client assets, associates them with abilities and phases, and aligns base and target events. Each research unit is an event asset; the canvas serializes assets rather than reproducing runtime timing.

We formulate completion of the unknown target assets from base audio, an optional prefix of completed target assets, a design description, and transformation hints. Our model fine-tunes Stable Audio 3~\cite{evans2026stable} with full base context and a partial target canvas. A signed \emph{soft mask} encodes both audio availability and each unknown event's requested change. Training hints use base--target distances in BEATs~\cite{chen2023beats} space, calibrated by training-set percentiles.

Our contributions are:
\begin{itemize}
\setlength{\itemsep}{2pt}
\setlength{\parsep}{0pt}
\item A collection, processing, and alignment pipeline for paired game-event SFX, applied to 5,633 target ability sequences.
\item A soft-mask-guided model that completes events from base audio, observed targets, and event-level transformation hints.
\item An empirical evaluation showing reconstruction gains over generic audio editors and the effectiveness of event-level transformation control, including coarse three-level hints.
\end{itemize}
Code will be released \href{https://github.com/Xinrui-Jiang/SfxSlotFill}{here}.

\begin{figure*}[t]
\centering
\includegraphics[width=\textwidth]{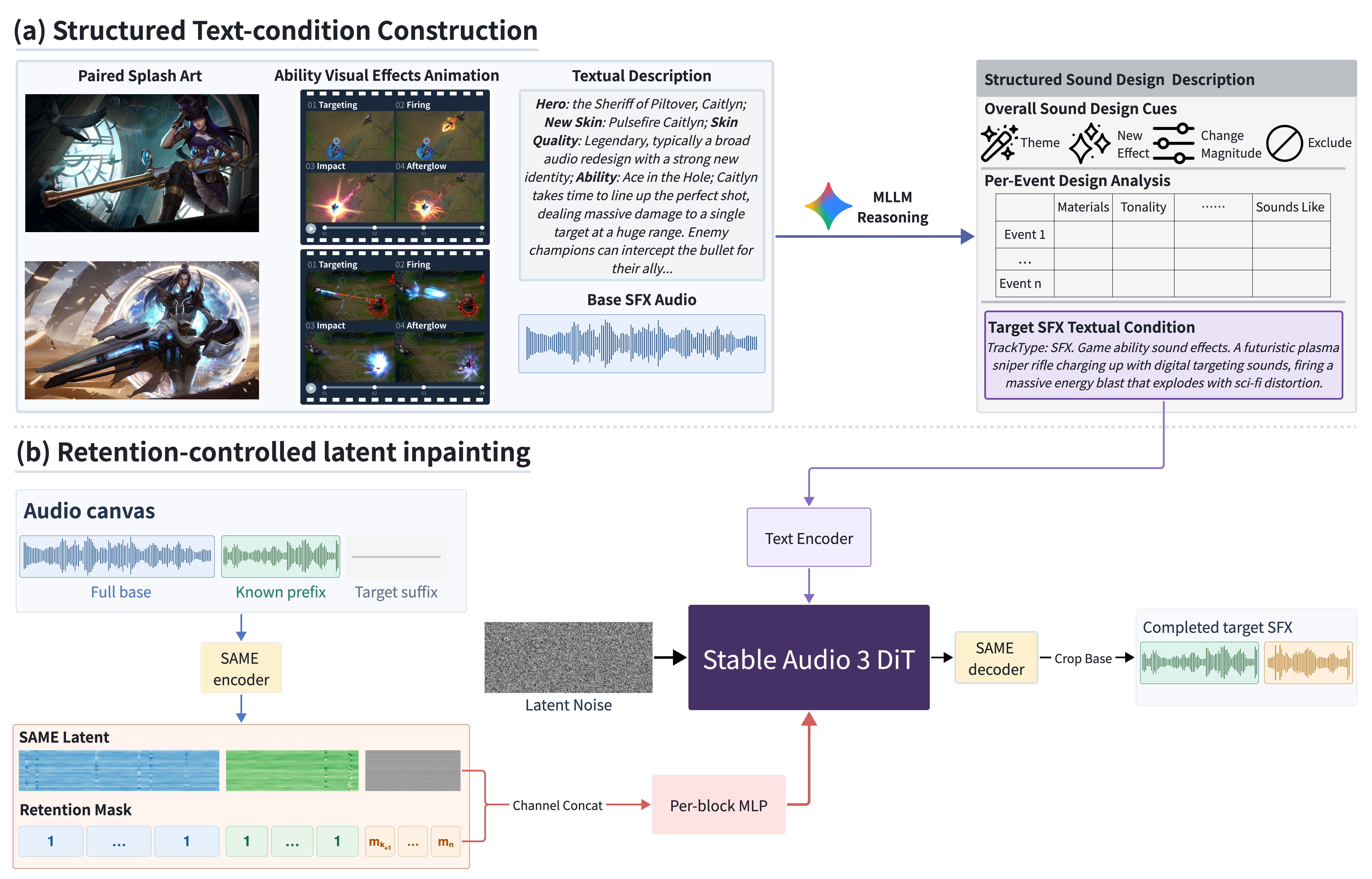}
\caption{Overview. \textbf{(a) Structured text-condition construction.} Gemini uses paired splash art, muted ability VFX clips, metadata, and base-skin SFX to produce a sound-design description. Skin-level cues and fixed-vocabulary event analyses are auxiliary reasoning; only the final one-sentence condition is used for model training and inference. Target-skin audio is never supplied to Gemini. \textbf{(b) Retention-controlled latent inpainting.} Full base audio, a known target prefix, and a silent target suffix form the waveform canvas, encoded by frozen SAME. Its latent $z$ is aligned with retention mask $M\in\mathbb{R}^{1\times T}$: $+1$ marks supplied audio; unknown event $j$ receives $m_j=-h_j\in[-1,0]$, expressing a requested change from retention ($0$) to full redesign ($-1$). Existing per-block MLPs project the channel-concatenated $[z;M]$ into every DiT block's residual stream; text enters through cross-attention. Sampling starts from Gaussian noise. After decoding and cropping to the target region, supplied target audio is reinserted unchanged. League of Legends artwork and gameplay imagery \textcopyright\ Riot Games, Inc.}
\label{fig:method}
\end{figure*}

\section{Related Work}
\label{sec:related}
\subsection{Audio generation and editing}
Text-guided audio editing must reconcile a requested change with preservation of the source. Pan et al.~\cite{pan2026survey} distinguish training-based adaptation from training-free use of foundation models. AudioLDM~\cite{liu2023audioldm} and Stable Audio 3~\cite{evans2026stable} supply generative priors. To learn the editing relationship, SAO-Instruct trains on free-form instructions~\cite{ungersboeck2025sao}, while SmartDJ decomposes requests into operations executed by a trained editor~\cite{lan2025smartdj}. Training-free editors reuse pretrained models: ZETA uses DDPM inversion~\cite{manor2024editing}, MelodyFlow adapts inversion to flow-matching music models~\cite{lelan2024melody}, and DirectAudioEdit uses inversion-free diffusion-prediction contrast~\cite{ge2026direct}.

Multimodal models broaden the available conditions. AudioX integrates text, visual, and audio inputs for generation~\cite{tian2025audiox}; Audio-Omni couples a frozen multimodal language model with a trainable DiT to unify generation and editing~\cite{tian2026omni}. These approaches make design intent more expressive, while control over which source characteristics should remain requires additional structure.

\subsection{Structured control for audio editing}
Structured conditions specify where and how audio may change. Stable Audio 3 inpainting supplies known audio and identifies regions to generate~\cite{evans2026stable}. AudioMorphix combines a time--frequency editing region with another recording as an acoustic reference~\cite{liang2025audiomorphix}; Audio ControlNet instead specifies attributes such as loudness, pitch, and event activity~\cite{zhu2026audiocontrolnet}. UNISON unifies generation and editing through task masks and channel-concatenated source-audio latents~\cite{li2026unison}, using explicit conditions to organize both tasks.

Cross-skin completion requires synthesizing missing events whose intended departures from the base differ. We combine full base audio and observed target assets in a shared canvas, and encode both availability and event-specific transformation magnitudes in a signed soft mask. This separates the region requiring synthesis from the requested degree of redesign.

\section{Data and Method}
\label{sec:method}
\subsection{Collection, processing, and alignment pipeline}
Unlike WavCaps audio--caption pairs~\cite{mei2023wavcaps}, our collection links base and target game-event recordings, grouped by ability.

Wwise banks from the game client are parsed with \texttt{wwiser} and rendered with \texttt{vgmstream}; CommunityDragon definitions provide ability, skin, visual-effect, and animation metadata. We extract audio from game-client version~16.13, with CommunityDragon metadata and supplementary audio banks from version~16.18. Naming conventions, visual-effect references, and animation cues associate sounds with abilities and phases. Skin-tag substitution and bank lookup establish base--target correspondences, including replacement, inheritance, silence, and additional events.

Events are ordered by phase and key. Base events define the skeleton, with target-only events inserted by phase. Audio is stored at 44.1\,kHz, without loudness normalization; continuous sounds are capped at 2\,s with a 50\,ms fade. Each sound is left-aligned in its window, with zero padding and 0.2\,s inter-window gaps. Window capacities use the maximum post-cap event duration across skins, replaced by the 95th percentile when the maximum exceeds twice the median, and rounded up to 1\,ms. Thus, corpus-wide duration metadata, including held-out skins, defines the layout; target-duration prediction is outside the task.

Filtering removes skins with at least 50\% inherited audible windows, base-identical sequences, silent-base cases, and duplicates, including held-out targets byte-identical to training targets. Splitting is by target skin, with shared base assets; it tests unseen skins, not necessarily unseen champions. The finalized collection contains 5,633 target ability sequences. Requiring sequences of at most 24\,s and an unknown event with a base counterpart yields 4,428 training, 259 validation, and 256 test sequences. Prefix enumeration yields 18,944 training examples per pass without crossing skin splits.

Gemini 3.1 Pro Preview generates a description of at most 40 words from base audio and its event schedule, base/target artwork, official descriptions, and muted ability videos when available. It never receives target audio. We prepend ``TrackType: SFX. Game ability sound effects.'' Skin-level design cues and per-event analyses over fixed vocabularies support the description construction (Fig.~\ref{fig:method}a). Only the final sentence conditions the model; the auxiliary analyses are not training inputs.

\subsection{Reference canvas and signed soft mask}
Let $B$ and $Y$ be aligned base and target canvases, $K$ the observed target events, and $U$ the remaining events. The model generates $Y_U$ conditioned on $B$, $Y_K$, text $c$, and hints $h_U$. In training, $K$ contains the first $k$ audible target events, for every $k$ from zero to one less than their total count. Zeroing unknown target events gives $Y_K^0$. The waveform reference and encoded training target are
\begin{equation}
 a=[B\mid 0_g\mid Y_K^0],\qquad x_0=E([B\mid 0_g\mid Y]),
\end{equation}
where $g=0.2$\,s and $E$ is the frozen SAME encoder. Its latents have 256 channels and a stride of 4,096 samples. The reference latent $E(a)$ is paired with a frame-aligned retention mask $M\in\mathbb{R}^{1\times T}$, where $T$ is the latent length:
\begin{equation}
 M_f=\begin{cases}
 +1,&\text{base audio or observed target events},\\
 -h_j,&\text{unknown target event }j,\\
 0,&\text{gaps and batch padding}.
 \end{cases}
 \label{eq:marker}
\end{equation}
The channel-concatenated $[E(a);M]$ is projected by the backbone's existing per-block MLPs and added to each DiT block's residual stream (Fig.~\ref{fig:method}b), without adding parameters. T5Gemma text embeddings enter through cross-attention. The mask communicates availability and requested change without enforcing waveform preservation. At $h_j=0$, its value coincides with padding; canvas context distinguishes events, and exact copying is not guaranteed.

\begin{table*}[t]
\centering
\caption{Test results with requested $k=2$ (256 sequences, 687 unknown windows). Prefix denotes observed target assets; oracle hints use unknown target audio. Best values per block are bold. SA3: Stable Audio 3; a2a: audio-to-audio; $\eta$: noise strength.}
\label{tab:main}
\small
\setlength{\tabcolsep}{7pt}
\begin{tabular}{lcccccc}
\toprule
Method & Prefix & Mel $\downarrow$ & MFCC $\downarrow$ & Loud $\downarrow$ & CLAP $\uparrow$ & Onset F1 $\uparrow$\\
\midrule
Copy-base & No & 12.43 & 73.9 & 7.00 & \textbf{0.661} & \textbf{0.601}\\
MelodyFlow~\cite{lelan2024melody} & No & 22.56 & 146.4 & 15.38 & 0.283 & 0.509\\
SA3 a2a ($\eta=0.7$) & No & 16.93 & 91.3 & 6.23 & 0.420 & 0.492\\
SA3 a2a ($\eta=0.5$) & No & 13.02 & 73.9 & 5.23 & 0.500 & 0.507\\
SA3 a2a + prefix & Yes & 15.37 & 95.2 & 7.91 & 0.492 & 0.485\\
ZETA~\cite{manor2024editing} & No & 15.60 & 99.4 & 8.80 & 0.504 & 0.517\\
DirectAudioEdit~\cite{ge2026direct} & No & 15.17 & 95.0 & 9.76 & 0.562 & 0.528\\
SmartDJ-Editor~\cite{lan2025smartdj} & No & 15.48 & 100.5 & 9.08 & 0.537 & 0.556\\
SAO-Instruct~\cite{ungersboeck2025sao} & No & 12.16 & 74.1 & 4.77 & 0.554 & 0.556\\
Ours, no-hint training & Yes & 9.86 & 58.9 & 4.47 & 0.653 & 0.534\\
Ours, constant $h=0.5$ & Yes & \textbf{9.39} & \textbf{55.2} & \textbf{3.72} & 0.660 & 0.546\\
\midrule
Ours, three-level oracle hints & Yes & 8.97 & 52.4 & 3.73 & 0.670 & \textbf{0.561}\\
Ours, continuous oracle hints & Yes & \textbf{8.93} & \textbf{52.1} & \textbf{3.69} & \textbf{0.673} & 0.556\\
\bottomrule
\end{tabular}
\end{table*}

\subsection{Calibrating transformation hints}
Let $e(\cdot)$ be the temporally averaged BEATs iter3+ AudioSet-2M embedding of a mono, peak-normalized asset padded or cropped to 5\,s. For a paired event,
\begin{equation}
 d_j=1-\cos(e(y_j),e(b_j)),\qquad h_j=F_{\rm train}(d_j),
 \label{eq:hint}
\end{equation}
where $F_{\rm train}$ is the empirical cumulative distribution over 18,641 audible training pairs, reused unchanged for validation and test. Percentile calibration spreads concentrated raw distances over $[0,1]$. We set $h_j=0$ for exact base copies and jointly empty pairs, and $h_j=1$ for unmatched audible or removed target events.

Training uses $y_j$ as supervision; deployment accepts $h_j$ as an input control. We evaluate continuous hints, quantization to $\{0,0.5,1\}$, constant $h_j=0.5$, and shuffling among unknown events within each sequence. Continuous and quantized hints are labeled \emph{oracle}, as they use unknown target audio.

\subsection{Training and inference}
We fine-tune the Stable Audio 3 medium-base DiT (1.4B parameters) with flow matching. Given target latent $x_0$ and Gaussian noise $\epsilon$, the interpolated input and target velocity are
\begin{equation}
 x_t=(1-t)x_0+t\epsilon,\qquad v^*=\epsilon-x_0.
\end{equation}
For predicted velocity $\hat v_f$ at frame $f$, the per-example loss is
\begin{equation}
 \mathcal L=\frac{\sum_f w_f\,\|\hat v_f-v^*_f\|_2^2}
 {C\sum_f w_f},\qquad C=256.
 \label{eq:loss}
\end{equation}
Here $C$ is the latent channel count; the weighted MSE averages over channels and frames. Weights are 1 on unknown-target audio, 0.25 on its within-window zero padding, and 0 elsewhere. All DiT parameters are fine-tuned for 40,000 updates using AdamW with learning rate $3\times10^{-5}$ and effective batch size 12.

Training replaces text by the empty string with probability 0.2. We use two-axis classifier-free guidance (CFG) over text and reference conditions $r=(E(a),M)$, comprising base audio, observed target events, and the retention mask. At the same $x_t,t$, let $v_{ij}$ denote the prediction with text present ($i=1$) or replaced by the empty-string embedding ($i=0$), and with $r$ present ($j=1$) or both its audio and mask channels zeroed ($j=0$). Text guidance is computed within each reference state, followed by reference guidance:
\begin{equation}
 \begin{aligned}
 u_R&=v_{01}+s_T(v_{11}-v_{01}),\\
 u_{\varnothing}&=v_{00}+s_T(v_{10}-v_{00}),\\
 v_{\rm CFG}&=u_{\varnothing}+s_R(u_R-u_{\varnothing}).
 \end{aligned}
 \label{eq:cfg}
\end{equation}
Here $s_T$ and $s_R$ set the text and reference guidance strengths, respectively. Validation selects $(s_T,s_R)=(2,1)$, yielding $v_{\rm CFG}=u_R=v_{01}+2(v_{11}-v_{01})$; no zero-reference prediction contributes at this setting. After Euler integration and SAME decoding, generated event windows are combined with the original observed target assets to form the output.

\section{Experiments}
\subsection{Protocol and comparisons}
Evaluation requests $k=2$ observed events, capped at $n_a-1$ for $n_a$ audible events. All 259 validation and 256 test sequences contribute 688 and 687 nonempty unknown windows, respectively. Scores are averaged within sequences and then across sequences, excluding silent targets and context. Reconstruction covers actual target durations, excluding excess generated tails.

We report 48-bin log-mel mean absolute error (Mel, dB), Euclidean distance between mean 20-dimensional MFCC vectors (MFCC), loudness error (Loud, dB), and LAION-CLAP audio--audio cosine similarity~\cite{wu2023clap}. Onset F1 uses \texttt{librosa} detection and \texttt{mir\_eval} scoring with a 0.1\,s tolerance~\cite{mcfee2015librosa,raffel2014mireval}. We also measure audio--text CLAP and target-to-generation PANNs KL~\cite{kong2020panns}.

Generic editors receive each base event and the same description. A sequence-level baseline inserts completed target assets into the base canvas and runs Stable Audio 3 audio-to-audio with noise strength 0.5 and eight sampling steps. Copy-base returns the base asset. No-hint training sets unknown-event markers to zero.

Validation sweeps select step and guidance settings. For continuous hints, 25 and 50 steps give test Mel 8.93, with CLAP 0.674 and 0.673, respectively. The prefix sweep uses 25 steps.

\subsection{Main results}
Our model achieves the lowest Mel, MFCC, and loudness errors and the highest CLAP similarity among all tested generative editors in Table~\ref{tab:main}. With event-level hints, it also exceeds copy-base on CLAP, 0.673 versus 0.661. Mel error is 26.6\% lower than SAO-Instruct, the strongest generic editor on this metric. The advantage extends to sequence-level Stable Audio 3 supplied with completed target assets.

Task adaptation and event-level control provide complementary benefits. No-hint training already surpasses the tested editors on Mel, MFCC, loudness, and CLAP, highlighting the value of training on paired game events. Constant hints further improve reconstruction, while assigning each event its own transformation level gives the strongest joint reconstruction and target agreement. The same progression holds on validation data. Audio--text CLAP and PANNs KL also improve, indicating closer semantic agreement. Copy-base preserves source onsets exactly; its highest onset F1 is consistent with the temporal correspondence between paired events.

\subsection{Hint ablations and prefix length}
Event correspondence is central to effective control. With the same checkpoint, continuous hints improve both Mel and CLAP over constant control in Table~\ref{tab:design}. Shuffling hint assignments removes this gain while preserving their value distribution, demonstrating the importance of matching each hint to its event. Three-level hints retain nearly all of the continuous-hint benefit, showing that coarse controls suffice without requiring precise transformation values.

\begin{table}[!ht]
\centering
\caption{Hint ablations on the full test set at $k=2$. All variants except the positive-marker model share one checkpoint.}
\label{tab:design}
\vspace{4pt}
\small
\setlength{\tabcolsep}{5pt}
\begin{tabular}{lcc}
\toprule
Hint setting & Mel $\downarrow$ & CLAP $\uparrow$\\
\midrule
Constant $h=0.5$ & 9.39 & 0.660\\
Shuffled percentile hints & 9.76 & 0.652\\
Three-level oracle hints & 8.97 & 0.670\\
Positive similarity markers & 9.29 & 0.670\\
Continuous oracle hints & \textbf{8.93} & \textbf{0.673}\\
\bottomrule
\end{tabular}
\end{table}

The signed percentile encoding outperforms positive similarity markers in both metrics, supporting its use for transformation control. Text guidance provides a complementary benefit: increasing $s_T$ from 1 to 2 raises validation CLAP from 0.664--0.669 to 0.680--0.681 across the reference-guidance sweep.

\begin{table}[!ht]
\centering
\caption{Prefix sweep on the same 111 test sequences with at least five audible events, using 25 steps and continuous oracle hints. Scored suffix events change with $k$.}
\label{tab:k}
\vspace{4pt}
\small
\setlength{\tabcolsep}{5pt}
\begin{tabular}{ccccc}
\toprule
& \multicolumn{2}{c}{Ours (oracle)} & \multicolumn{2}{c}{Copy-base}\\
$k$ & Mel $\downarrow$ & CLAP $\uparrow$ & Mel $\downarrow$ & CLAP $\uparrow$\\
\midrule
0 & 9.34 & 0.661 & 12.00 & 0.668\\
1 & 9.10 & 0.668 & 12.37 & 0.663\\
2 & 8.75 & 0.671 & 12.65 & 0.659\\
3 & 8.57 & 0.675 & 13.15 & 0.657\\
4 & 9.01 & 0.666 & 13.61 & 0.649\\
\bottomrule
\end{tabular}
\end{table}

Our model maintains lower Mel error than copy-base at every prefix length in Table~\ref{tab:k}. On all 256 test sequences at $k=0$, it achieves Mel 9.35, compared with 11.94 for copy-base and 14.48 for Stable Audio 3 audio-to-audio. The reconstruction advantage therefore persists without any completed target audio, supporting completion from base assets and design controls alone.

\begin{samepage}
\noindent\textbf{Evaluation scope.} Continuous and quantized hints use target-derived oracle information; designer-specified hints remain to be validated. The objective metrics measure acoustic agreement, while production usability calls for designer listening studies~\cite{desbos2026production}. Evaluation also uses corpus-wide duration metadata, motivating tests with independently specified event layouts.\par
\end{samepage}

\section{Conclusion}
We developed an event-alignment pipeline and a retention-controlled model for game SFX completion. Paired event data enables effective task adaptation, while a signed mask exposes event-level transformation controls. Experiments demonstrate reconstruction gains over general-purpose editors and the value of correctly assigned hints, with three levels retaining most of the continuous-hint benefit.

Expanding paired data across games is the next priority. This requires adapting asset extraction and alignment to different audio systems while preserving correspondence among gameplay events. Larger collections with more event types and design variants would enable studies of data scaling and transfer between games.

\section{Acknowledgments}
ChatGPT polished the writing. Claude Code cross-checked experiment code and batch results. Gemini 3.1 Pro Preview generated the textual conditions in Section~3.1.
\clearpage
\begingroup
\def\baselinestretch{1}
\fontsize{10}{12}\selectfont
\balance
\bibliographystyle{IEEEbib}
\bibliography{paper_refs}
\endgroup
\end{document}